\documentclass[aps,prx,reprint,nofootinbib,preprintnumbers,amsmath,amssymb,amsfonts]{revtex4-2}
\pdfoutput=1
\usepackage[utf8]{inputenc}
\usepackage{graphicx}
\usepackage{hyperref}

\newcommand{\cF}{\mathcal{F}}

\newcommand{\LQG}{\Lambda_{\rm QG}}

\newcommand{\lQG}{\lambda_{\rm QG}}

\begin{document}

\title{Persistence of the Pattern in the Interior of 5d Moduli Spaces}

\author{Tom Rudelius}
\email{thomas.w.rudelius@durham.ac.uk}
\affiliation{Department of Mathematical Sciences, Durham University, Durham DH1 3LE United Kingdom}

\date{\today}

\begin{abstract}
Castellano, Ruiz, and Valenzuela recently observed a remarkable ``pattern'' in infinite-distance limits of moduli spaces in quantum gravity, which relates the field space variation of the mass of the lightest tower of particles to the field space variation of the species scale.
In this work, we show how a version of this pattern can be proven to hold for BPS particles and strings throughout the vector multiplet moduli space of a 5d supergravity theory, even in regions where the particle masses and string tensions are substantially modified relative to their asymptotic behavior in the infinite-distance limits. This suggests that a suitably defined version of the pattern may hold not merely in the asymptotic limits of moduli space, but in the interior as well.
\end{abstract}

% insert suggested PACS numbers in braces on next line
% insert suggested keywords - APS authors don't need to do this
%\keywords{}

\maketitle

\section{Introduction}\label{INTRO}

Recent explorations of moduli spaces in quantum gravity have uncovered evidence of universal structures within infinite-distance limits. As pointed out long ago
by Ooguri and Vafa \cite{Ooguri:2006in}, such limits seem to feature a tower of light particles, whose masses decay exponentially with geodesic distance $||\phi||$ as $m \sim \exp(- \alpha ||\phi||)$. Such a tower of light particles implies that quantum gravity becomes strongly coupled at an energy scale $\LQG$ (often referred to as the ``species scale'') which is parametrically smaller than the Planck scale, and which itself decays as $\LQG \sim \exp(-\lambda_{\rm QG} ||\phi||)$.

It has long been understood that these exponential decay coefficients $\alpha$, $\lambda_{\rm QG}$ should be order-one numbers in Planck units,\footnote{We will set $8 \pi G = M_{\textrm{Pl};d}^{2-d} =1$ throughout this letter.} but recently these values have been pinned down with greater precision. Reference \cite{Etheredge:2022opl} presented strong evidence that the coefficient $\alpha$ of the lightest tower in a given infinite-distance limit must satisfy $\frac{1}{\sqrt{d-2}} \leq \alpha \leq \sqrt{\frac{d-1}{d-2}}$ in $d$ spacetime dimensions, and several subsequent works \cite{vandeHeisteeg:2023ubh, Calderon-Infante:2023ler, vandeHeisteeg:2023uxj} argued that the species scale should decay with a coefficient $\frac{1}{\sqrt{(d-1)(d-2)}} \leq \lQG \leq \frac{1}{\sqrt{d-2}}$.

Finally, \cite{Castellano:2023stg,Castellano:2023jjt} provided strong evidence for not merely an inequality, but actually an \emph{equality}, which relates the two decay coefficients mentioned above. In the case of a 1-dimensional moduli space, this equality takes the form $\alpha \lQG = 1/(d-2)$. More generally, the equality takes the form
\begin{equation}
    \frac{\vec{\nabla} m}{m} \cdot  \frac{\vec{\nabla} \LQG}{\LQG} = \frac{1}{d-2}\,,
    \label{patterneq}
\end{equation}
where $m$ is the characteristic mass scale of the lightest tower in the infinite distance limit, the gradient $\vec\nabla$ involves the derivative with respect to every massless modulus of the theory, and the inner product is taken with respect to the metric $g_{ij}$ on moduli space, which is parametrized by vacuum expectation values of the massless scalar fields. Plugging in $m = m_0 \exp(-\alpha \phi)$ and $\LQG = \Lambda_0 \exp(-\lQG \phi)$, we see that \eqref{patterneq} indeed reduces to $\alpha \lQG = 1/(d-2)$ for a 1-dimensional moduli space.

In \cite{Castellano:2023stg,Castellano:2023jjt}, equation \eqref{patterneq}--which the authors referred to as the ``pattern''--was conjectured to apply in any infinite-distance limit in the moduli space. Within such limits, the pattern may be viewed as a consequence of the Emergent String Conjecture \cite{Lee:2019wij}, which leads to tight constraints on the set of exponentially light towers and the relationships between them \cite{taxonomy}. The Emergent String Conjecture holds that every infinite-distance limit is either a decompactification limit or an emergent string limit in an appropriate duality frame, which in turn implies that the lightest tower of particles is either a tower of Kaluza-Klein modes or a tower of string oscillation modes. In addition, it suggests that the scale $\LQG$ should be (roughly) identified with either a string scale or a higher-dimensional Planck scale, which can be used to justify the pattern \eqref{patterneq}, as well as many of the other bounds on light towers and the species scale mentioned above \cite{taxonomy}.

It is not so clear, however, how this whole story carries over to the interior of moduli space, which is the region of greatest practical importance for phenomenology and cosmology.
As pointed out in \cite{Castellano:2023jjt}, even the notion of the ``lightest tower'' in moduli space is not generically well-defined outside the asymptotic regimes of moduli space. One exception to this, however, comes from towers of BPS particles, which remain well-defined throughout moduli space.
Similarly, although the definition of the species scale may be subtle, the tension $T$ of a BPS string is well-defined throughout moduli space, and from it we may define a string scale via $M_{\rm string
} \equiv \sqrt{2 \pi T}$.

One setting in which both BPS particles and BPS strings appear is supergravity in five dimensions, which offers a rich and diverse landscape of quantum gravity theories arising from M-theory compactifications on Calabi-Yau threefolds.
In this work, we explore this landscape, demonstrating that a suitably refined version of \eqref{patterneq} persists even into the interior of moduli space, where other universal features of infinite-distance limits cease to be valid.

In particular, we focus on vector multiplet moduli spaces of 5d supergravity theories that feature emergent string limits, in which a BPS string becomes tensionless in the asymptotic limit. In such theories, we find that the pattern in 
\eqref{patterneq} is satisfied throughout vector multiplet moduli space provided that (a) $m$ is defined to be the mass scale associated with a tower of BPS particles which become massless in some infinite-distance limit, and (b) the quantum gravity scale $\LQG$ is defined to be the string scale $M_{\rm string
} = \sqrt{2 \pi T}$ of an emergent BPS string of tension $T$. Remarkably, we find that \eqref{patterneq} remains satisfied even when the lengths of the individual vectors ${\vec\nabla m }/{ m}$ and ${\vec\nabla \LQG }/{ \LQG}$ vary significantly throughout moduli space, and it remains satisfied even after passing (via flop transitions of the Calabi-Yau threefold) into distinct phases of the theory.

Our analysis suggests that the pattern is indeed pointing toward a universal feature of quantum gravity, and it further suggests a possible refinement of the pattern. Namely, our results suggest that one should set $m$ in \eqref{patterneq} to be the mass scale associated with the lightest tower of Kaluza-Klein modes or string oscillation modes (which become massless in an appropriate infinite-distance limit), and one should take the quantum gravity scale $\LQG$ to be a string scale or higher-dimensional Planck scale (each of which also vanish in an infinite-distance limit). This choice for the quantum gravity scale $\LQG$ agrees with the species scale in asymptotic limits, but in the interior of moduli space the two may differ in subtle but important ways, as we discuss in \S\ref{CONC} below. Although the full scope of this refinement is unclear at present, it is encouraging that some features of the asymptotic limits of moduli space seem to apply in the interior as well. Clearly, further research is needed.

The remainder of this letter is structured as follows: in \S\ref{5DSUGRA}, we review relevant aspects of 5d supergravity. In \S\ref{PATTERN}, we show how equation \eqref{patterneq}, when properly defined, persists throughout vector multiplet moduli spaces of 5d supergravity theories. In \S\ref{CONC}, we end with a discussion of implications and possible refinements of the pattern.

\section{Basics of 5d Supergravity}\label{5DSUGRA}

In this section, we review relevant aspects of 5d supergravity. For further details, see \cite{Alim:2021vhs}.

Many features of a 5d supergravity theory are captured by its prepotential, a cubic function of the coordinates $Y^I$:
\begin{equation}
\mathcal{F} = \frac{1}{6} C_{I J K} Y^I Y^J Y^K \,,
\end{equation}
where here and henceforth repeated indices are summed.
In an M-theory compactification to 5d on a Calabi-Yau threefold $X$, indices $I,J,K$ run from $0$ to $h^{1,1}(X)-1$, the constants $C_{IJK}$ are the triple intersection numbers of the manifold, and the moduli $Y^I$ are volumes of certain two-cycles.
The vector multiplet moduli space is given by the slice $\cF = 1$, which means that this moduli space has dimension $n \equiv h^{1,1}(X)-1$. 

At a generic point in moduli space, the gauge group is $U(1)^{h^{1,1}(X)}$, and the gauge kinetic matrix is given by
\begin{equation}
  a_{I J} = \mathcal{F}_I \mathcal{F}_J - \mathcal{F}_{I J} ,
  \label{eq:gaugekinetic}
\end{equation}
with
\begin{equation}
   \mathcal{F}_I = \partial_I \cF= \frac{1}{2} C_{I J K} Y^J Y^K , 
   ~~ \mathcal{F}_{I
   J} = \partial_I \partial_J \cF  = C_{I J K} Y^K .
   \label{eq:prepottrip}
   \end{equation}
The metric on moduli space is given (up to a factor of $1/2$) by the pullback of $a_{IJ}$ to the $\cF =1$ slice:\footnote{Note that our definition of the moduli space metric $g_{ij}$ differs from that of \cite{Alim:2021vhs} by a factor of $2$: $g_{ij}^{\rm here} =  \frac{1}{2} \mathfrak{g}_{ij}^{\rm there}$.}
\begin{equation}
g_{ij} =  \frac{1}{2} a_{IJ} \partial_i Y^I \partial_j Y^J,
\label{hgmet}
\end{equation}
where $\partial_i \equiv \frac{\partial}{\partial \phi^i}$, and we have parametrized $Y^I = Y^I(\phi)$ in terms of the $n$ moduli $\phi^i$, $i = 1,...,n$.

Using the definitions above, it is possible to prove the following useful relation between the inverse metric $g^{ij}$ and the inverse gauge kinetic matrix $a^{IJ}$ \cite{Alim:2021vhs}:
\begin{align}
a^{IJ} =  \frac{1}{2} g^{ij} \partial_i Y^I \partial_j Y^J + \frac{1}{3} Y^I Y^J,
\label{inverseidentity}
\end{align}
Similarly, using the simple identities $Y^J \cF_{IJ} = 2 \cF_I$, $Y^I \cF_I = 3 \cF = 1$, and $a_{IJ}a^{JK}= \delta_I^K$, it is easy to see that
\begin{equation}
    a_{IJ} Y^J = \cF_I ~~\Rightarrow~~ a^{IJ} \cF_{I} = Y^J\,.
    \label{YandFidentity}
\end{equation}

A particle is labeled by a vector of electric charges, $q_I$. The mass of such a particle is bounded from below by the BPS bound:
\begin{align}
m(q_I) \geq  \left( \sqrt{2} \pi\right)^{1/3} |Z| =  \left( \sqrt{2} \pi \right)^{1/3} |q_I Y^I|\,,
\label{BPSbound}
\end{align}
Particles that saturate the BPS bound are called BPS particles.

Meanwhile, a string may carry magnetic charge under the gauge fields, which is labeled by a magnetic charge vector $\tilde{q}^I$. The tension of such a charged string is bounded from below by the BPS bound for strings:
\begin{align}
T(\tilde q^I) \geq  \frac{1}{2} \left( \frac{1}{\sqrt{2} \pi}\right)^{1/3} |\tilde Z| = \frac{1}{2} \left( \frac{1}{\sqrt{2} \pi}\right)^{1/3}  |\tilde q^I \mathcal{F}_I|\,.
\label{BPSstringbound}
\end{align}
Strings that saturate this bound are called BPS strings.

\section{The Pattern in 5d Supergravity}\label{PATTERN}

Suppose there exists an infinite-distance limit in vector multiplet moduli space in which a tower of BPS particles of mass $m_k = k m$ become light, and a BPS string of tension $T$ becomes tensionless. The results of \cite{Etheredge:2022opl, Rudelius:2023odg, taxonomy} indicate that in the asymptotic limit, the tension of the BPS string and the mass of the BPS particles will satisfy
\begin{align}
g^{ij} \frac{\partial_i m}{m} \frac{\partial_j m}{m} = \frac{4}{3}\,,~~~~~g^{ij} \frac{\partial_i T}{T} \frac{\partial_j T}{T} = \frac{4}{3}\,,
\label{KKandstringlength}
\end{align}
and
\begin{align}
g^{ij} \frac{\partial_i m}{m} \frac{\partial_i T}{T} = \frac{2}{3} \,.
\label{5dpattern}
\end{align}
Here, \eqref{5dpattern} matches the pattern equation \eqref{patterneq} after setting $d=5$, $\LQG = M_{\rm string} \equiv \sqrt{2 \pi T}$. In what follows, we will prove that \eqref{5dpattern} remains true at \emph{all} points of the vector multiplet moduli space, even points located in different phases of the theory. This is remarkable because the relations in \eqref{KKandstringlength} do not, in general, hold beyond the asymptotic regimes of moduli space.

To prove this, we must use one important fact: the tower of light particles (of charge $k q_I$, for increasing $k$) and the asymptotically tensionless BPS string (of magnetic charge $\tilde q^I$) have vanishing Dirac pairing, i.e., $q_I \tilde q^I = 0$. This fact follows from the discussion in \S5 of \cite{Alim:2021vhs}, which showed that all asymptotic limits of the moduli space feature a tensionless BPS string and a tower of light BPS particles with vanishing Dirac pairing.\footnote{At least, any such limit features a tensionless BPS string of charge $\tilde q^I$ and a set of charges $k q_I, k \in \mathbb{Z}$ with $q_I \tilde q^I = 0$ and $Z = q_I Y^I \rightarrow 0$. That these charges are occupied by BPS particles remains to be proven in full generality, but this is true in all known examples and follows from many well-supported quantum gravity conjectures \cite{Alim:2021vhs}.} In contrast, boundaries of moduli space involving light particles and tensionless strings of nontrivial Dirac pairing ($q_I \tilde q^I \neq 0$) lie at finite distance rather than infinite distance.\footnote{Geometrically, such finite-distance boundaries correspond to the collapse of a divisor to a point, whereas infinite-distance boundaries correspond to the collapse of the entire Calabi-Yau threefold \cite{Witten:1996qb}.}

With this fact in hand, the relation \eqref{5dpattern} follows straightforwardly from the definitions and identities of \S\ref{5DSUGRA}. To begin, we plug the formulas for the mass (tension) of BPS particles (strings) into the left-hand side of \eqref{5dpattern}:
\begin{align}
g^{ij} \frac{\partial_i m}{m} \frac{\partial_j T}{T} = \frac{\partial_i (q_I Y^I)}{q_K Y^K}  \frac{\partial_j (\tilde q^J \cF_J)}{\tilde q^L \cF_L}\,.
\end{align}
From here, we use the definition of $\cF_{IJ}$ in \eqref{eq:prepottrip} to set
$\partial_j \cF_J = \partial_j Y^P \cF_{PJ}$, and we use the identity of \eqref{inverseidentity} to rewrite this as
\begin{align}
g^{ij} \frac{\partial_i m}{m} \frac{\partial_j T}{T} &=  \frac{ 2 q_I a^{IP} \cF_{PJ} \tilde q^J}{(q_K Y^K)(\tilde q^L \cF_L)} - \frac{2}{3} \frac{(q_I Y^I) \cF_{PJ} Y^P \tilde q^J }{(q_K Y^K)(\tilde q^L \cF_L)}
\nonumber \\
&=  \frac{2 q_I a^{IP} \cF_{PJ} \tilde q^J}{(q_K Y^K)(\tilde q^L \cF_L)} - \frac{4}{3}\,,
\label{leftoff}
\end{align}
where in the second line we have used the fact that $\cF_{PJ} Y^P = 2 \cF_J$.
Next, by the definition of $a_{IJ}$ in \eqref{eq:gaugekinetic}, we have
\begin{align}
\delta^I_J = a^{IP}a_{PJ} = a^{IP} (\cF_P \cF_J - \cF_{PJ})\,,
\end{align}
which implies
\begin{align}
a^{IP}\cF_{PJ}= - \delta^I_J + a^{IP} \cF_P \cF_J = - \delta^I_J + Y^I \cF_J \,.
\end{align}
where in the last equation we have used the identity in \eqref{YandFidentity}. Plugging this into \eqref{leftoff}, we arrive finally at
\begin{align}
g^{ij} \frac{\partial_i m}{m} \frac{\partial_j T}{T} &= \frac{2}{3} - \frac{2 q_I \tilde q^I}{(q_K Y^K)(\tilde q^L \cF_L)}\,.
\end{align}
Imposing the vanishing of the Dirac pairing between the BPS particles and the BPS string, $q_I \tilde q^I = 0$, we find the desired result \eqref{5dpattern}.

\section{Discussion}\label{CONC}

Within 5d supergravity, we have shown that \emph{any} BPS particle of charge $q_I$ and \emph{any} BPS string of charge $\tilde q^I$ with vanishing Dirac pairing $q_I \tilde q^I = 0$ satisfy the relation \eqref{5dpattern}. This implies the pattern of \eqref{patterneq} after setting $\LQG = M_{\rm string} = \sqrt{2 \pi T}$. It is relatively well-established that this relation is satisfied in the asymptotic limits of moduli space; our result shows that it extends to the \emph{entire} moduli space. This agrees with observations made in \cite{Castellano:2023jjt}, which showed that the pattern is satisfied exactly (i.e., not merely at leading order in an expansion) in certain infinite-distance limits in vector multiplet moduli spaces of 4d supergravity theories.

Our result is nontrivial for a couple of reasons. To begin, it is not hard to find examples in which the closely related formulas of \eqref{KKandstringlength} are badly violated in the interior of moduli space, even though they are indeed satisfied in the asymptotic limits.\footnote{Type I' string theory in 9d is another well-studied case where formulas analogous to those of \eqref{KKandstringlength} are violated--see \cite{Etheredge:2023odp} for further details.} In some cases, the length of the vector $\partial_i m / m$ may even diverge in certain limits of moduli space, yet even then the relation \eqref{5dpattern} remains valid. (The $h^{1,1} = 3$ geometry of \cite{Klemm:1996hh}, studied extensively in \cite{Alim:2021vhs}, is one example where this occurs.) Secondly, the interior of a 5d vector multiplet moduli space can feature complicated phase transitions (geometrically realized by flop transitions or Weyl reflections \cite{Gendler:2022ztv}), each of which involves a modification of the triple intersection numbers $C_{IJK}$ and hence of the prepotential $\cF$. Remarkably, our derivation of \eqref{5dpattern} is unaffected by such modifications, and thus the relation persists even across the distinct phases of vector multiplet moduli space. This suggests that, whereas many features of infinite-distance limits of moduli spaces in quantum gravity may be consequences of the weakly coupled lamppost, the pattern of \eqref{patterneq}, suitably interpreted, may have a wider range of applicability.

In addition, our work suggests a possible modification of the pattern of \eqref{patterneq}, in which the species scale is replaced by a string scale or higher-dimensional Planck scale. One possibility takes the form
\begin{align}
   \frac{\vec{\nabla} m}{m} \cdot  \frac{\vec{\nabla} M}{M} &= \frac{1}{d-2}\,,\nonumber \\
 \text{with} ~ m = \min(\mathcal{T}) ~  &\text{for some} ~ M  \in \mathcal{S} \cup \mathcal{P} ,
    \label{revpatterneq}
\end{align}
where $\mathcal{T}$ is the set comprised of the mass scales of all of the Kaluza-Klein towers or string oscillator towers which become light in the various asymptotic limits of the theory, $\mathcal{S}$ is the set of all of the string scales $M_{\rm string} = \sqrt{2 \pi T}$ of the fundamental strings which become light in the emergent string limits of the theory, and $\mathcal{P}$ is the set of higher-dimensional Planck scales that arise in the decompactification limits of the theory. In many cases, including in asymptotic limits of moduli space, $M$ can in fact be chosen to be the smallest mass scale in the set $\mathcal{S} \cup \mathcal{P}$.

This modified version of the pattern is more or less equivalent to \eqref{patterneq} in infinite-distance limits of moduli space. The advantage of the modification can be seen, for instance, in Type IIB string theory in 10 dimensions. Here, the species scale $\Lambda_{\rm sp
}$ (defined as the coefficient of the $R^4$ term in the action) varies smoothly over moduli space, reaching a critical point with $\vec \nabla \Lambda_{\rm sp
} = 0$ at $\tau = i$ \cite{vandeHeisteeg:2023dlw}. As a result, \eqref{patterneq} breaks down in a neighborhood of this point if the quantum gravity scale is simply defined to be the species scale, $\LQG = \Lambda_{\rm sp
}$, and $m$ is simply defined to be the mass scale of the lightest tower. A similar breakdown may be expected to occur near critical points of the species scale in 5d supergravity as well, which can occur at a boundary between two phases of the theory. (The symmetric flop geometry of \cite{Alim:2021vhs} is one example where this occurs.)

In contrast, the choice of the scale $M$ used in \eqref{revpatterneq} has the advantage that it, like the mass scale of the lightest tower $m$, can have discontinuous first derivative across the boundaries between different regions of moduli space, as different string scales/higher-dimensional Planck scales may satisfy the bound at different points in moduli space. As a result, \eqref{revpatterneq} is satisfied at all points in Type IIB moduli space and all points in the vector multiplet moduli space of the aforementioned symmetric flop example of \cite{Alim:2021vhs}.

The full scope of the relation proposed in \eqref{revpatterneq} remains to be understood. Indeed, it is not even clear that it holds in 5d supergravity theories when the lightest tower consists of non-BPS particles (such as the Kaluza-Klein modes for a decompactification limit to 11d M-theory).
Nonetheless, it is exciting to think that the pattern first observed in \cite{Castellano:2023stg, Castellano:2023jjt} may have a broader range of applicability than previously realized, and it is refreshing to see that some of the conjectured features of quantum gravity may extend beyond the infinite-distance lamppost.

%\begin{acknowledgments}
\vspace*{1cm}
{\bf Acknowledgments.} We thank Muldrow Etheredge,  Ben Heidenreich, Damian van de Heisteeg, Jacob McNamara, Ignacio Ruiz, Cumrun Vafa, Irene Valenzuela, and Max Wiesner for useful discussions, and we further thank Ben Heidenreich, Ignacio Ruiz, and Irene Valenzuela for comments on a draft.
This work was supported in part by STFC through grant ST/T000708/1.

\bibliography{ref}

\end{document}